\documentclass[usegraphicx,usenatbib,onecolumn,doublespace]{mn2e}
\usepackage{bm}
\usepackage{amsmath,amssymb}

\begin{document}
\title{Primordial magnetic fields with X-ray and S-Z cluster survey
} 
\author[Tashiro, H. et al.]
{Hiroyuki Tashiro$^{1,2}$, Keitaro Takahashi$^{3,4}$, Kiyotomo Ichiki$^{4}$\\
$^1$ Centre for Cosmology, Particle Physics and Phenomenology (CP3),
Universit\'{e} catholique de Louvain,%
B-1348 Louvain-la-Neuve, Belgium;\\
$^2$ Department of Physics, Arizona State University, Tempe, 
AZ 85287-1404, USA\\
$^3$ Faculty of Science, Kumamoto University, 2-39-1, Kurokami, Kumamoto 860-8555, Japan\\
$^4$ Department of Physics and Astrophysics,
Nagoya University, Chikusa-ku Nagoya 464-8602 Japan
}
\date{\today}
\maketitle

\begin{abstract}
The effect of primordial magnetic fields on X-ray or S-Z galaxy
cluster survey is investigated. After recombination, the primordial
magnetic fields generate additional density fluctuations. Such
density fluctuations enhance the number of galaxy clusters.
Taking into account the density fluctuations generated by primordial
magnetic fields, we calculate the number of galaxy clusters based
on the Press-Schechter formalism.
Comparing with the results of Chandra X-ray galaxy cluster survey,
we found that the existence of primordial magnetic fields with
amplitude larger than 10 nGuass would be inconsistent.
Moreover, we show that S-Z cluster surveys also have a sensitivity
to constrain primordial magnetic fields. Especially SPT S-Z cluster
survey has a potential to constrain the primordial magnetic fields
with several nano Gauss.
\end{abstract}

\section{introduction}
The origin of large-scale magnetic fields observed
in galaxies and galaxy clusters still remains unclear.
The most widely accepted theory is the astrophysical scenario
in which seed magnetic fields are generated by the battery 
mechanism in astrophysical phenomena and amplified by
the dynamo mechanism in the interstellar or intergalactic medium
\citep{barndenburg-subramanian-2005}.
However, there is uncertainty about the efficiency of the
dynamo mechanism \citep{kulsrud-cowley-1997,giovannini-2004}.
The recent studies on Faraday rotation measurements of high 
redshift quasars suggest the existence of the $\mu$Gauss 
magnetic fields in high-redshift galaxies
\citep{kronberg-benet-2008,bernet-miniati-2008}.
The existence of such magnetic fields may also challenge the dynamo 
scenario.

Another candidate for the origin of the galactic magnetic
fields is primordial magnetic fields which is generated in the
early universe \citep{widrow-2002,2005PhRvL..95l1301T,
2006Sci...311..827I,giovannini-2008}. If primordial
magnetic fields existed, Big Bang Nucleosynthesis (BBN) 
and Cosmic Microwave Background (CMB)
anisotropy suffer the effect of primordial magnetic fields.
Therefore, the constraint on the primordial magnetic fields through
BBN and CMB anisotropy has been studied by many authors.

After the recombination epoch, primordial magnetic fields generate
additional density fluctuations by the Lorentz force
\citep{wasserman-78,kim-olinto-96} and many authors have studied
their effects on the evolution of the large scale
structures; the redshift-space matter power spectrum,
the epoch of reionization, and 21 cm fluctuations
\citep{gopla-sethi-2003,
sethi-subramanian-2005,tashiro-sugiyama-reionization,
tashiro-sugiyama-21cm,schleicher-banerjee-2009,
sethi-subramanian-2009,2010arXiv1006.4242S}. 
It was found that magnetic fields as small as
a few nano Gauss can give strong cosmological impacts.
Therefore detailed observations planned in near future have
the potential to set further constraints on primordial magnetic fields.

In this paper, we study the effect of primordial magnetic
fields on the mass function of galaxy clusters.
The additional density fluctuations generated by the primordial magnetic
fields enhance the formation of galaxy clusters.
Thus, number count of clusters can constrain their amplitude
as well as the standard cosmological parameters such as the amplitude
of density fluctuations ($\sigma_8$) and energy density of matter
($\Omega_{\rm m}$). Specifically we investigate the potential of
the cluster number count by X-ray observations and Sunyaev-Zel'divich
(S-Z) survey.

We can find massive clusters by the observation of the X-ray emitted
from the hot intracluster gas.
X-ray-flux-selected cluster samples with the calibration between
the X-ray temperature and the cluster mass give the mass function 
of galaxy clusters.
\citet{vikhlinin-cosmology-2009} and \citet{mantz-allen-2010} 
applied it to give a constraint on the cosmological 
parameters such as the equation of state of dark energy.

The S-Z effect is the scattering of CMB photons 
by the hot intracluster electron gas \citep{szeffect-1972} and
is also a powerful tool for detecting galaxy clusters at high redshifts.
Combining the S-Z galaxy survey with X-ray or optical observations,
we can obtain the mass- or redshift-abundance of the number of
galaxy clusters. There are many observation projects
carried out or planned. In particular, {\it Planck} and the South
Pole Telescope (SPT) is expected to give large catalogs of S-Z
galaxy clusters and the significant constraints on cosmological
parameters. 

The rest of the paper is organized as follows. In Sec.~II,
we give a description of the density fluctuation generation
by primordial magnetic fields after the recombination epoch.
In Sec.~III, we study the effect of primordial magnetic fields
on the mass function by using the Press-Schechter formalism 
calibrated to numerical simulations. We also compare our result with
the mass function derived from {\it Chandra} observations and obtain
a constraint on the strength of primordial magnetic fields.
In Sec.~IV, we calculate the number count for S-Z galaxy cluster
surveys and discuss the potential of {\it Planck} and SPT
to give constraints on the primordial magnetic fields. 
We conclude in Sec. V.
Through this paper, we assume a $\Lambda$CDM cosmological model with
$h=0.72$, $\Omega_{\rm m}=0.3$, $\Omega_{\rm b}=0.05$ and $\sigma_8=0.75$.

\section{density fluctuations due to primordial magnetic fields}\label{density}

In this section, we calculate the density fluctuations produced by
primordial magnetic fields. First, we make some assumptions about
primordial magnetic fields. Because our interest is in relatively large
length scales, we can assume that the back-reaction of the fluid velocity
to magnetic fields is small. Therefore, we consider the case where
primordial magnetic fields are frozen in cosmic baryon fluids, 
\begin{equation}
{\bm B} (t,{\bm x}) ={ {\bm B} _0({\bm x}) \over a^2(t)}.
\end{equation}
Here $B_0({\bm x})$ is the comoving strength of magnetic fields and
$a(t)$ is the scale factor which is normalized as $a(t_0)=1$
at the present time, $t_0$. For simplicity, we assume that primordial
magnetic fields are statistically homogeneous and isotropic and
have the power-law spectrum with the power-law index $n$,
\begin{equation}
\langle B_{0i}({\bm k_1} ) B_{0j} ^* ({\bm k_2} )\rangle = 
{(2 \pi)^3 \over 2}
\delta ({\bm k_1}-{\bm k_2}) \left(\delta_{ij}-{k_{1i} k_{2j} \over k_1^2 } \right) P_B(k) ,
\nonumber
\end{equation}
\begin{equation}
P_B(k)=B^2 _{\rm n}
\left ({k \over k_{\rm n}} \right)^{n},
\label{eq:powerlaw}
\end{equation}
where $\langle ~ \rangle$ denotes the ensemble average,  
$B_{0i}({\bm k} )$ are Fourier components of $B_{0i}({\bm x})$, $k_{\rm n}$ 
is the wave number of an arbitrary normalized scale and 
$B_{\rm n}$ is the magnetic field strength at $k_{\rm n}$.

Our interest is to constrain the magnetic field strength on a certain scale
in the real space. Therefore, we have to convolve the power spectrum with 
a Gaussian filter transformation of a comoving radius $\lambda$,
in order to get the magnetic field strength in the real space
\citep{mack-kahniashvili-02},
\begin{equation}
B_\lambda ^2 \equiv \langle B_{0i}({\bm x} ) B_{0i}  ({\bm x} )\rangle |_\lambda =
{1 \over (2 \pi)^3} \int d^3 k B^2 _{\rm n}
\left ({k \over k_{\rm n}} \right)^{n} 
\left | \exp \left(- {\lambda ^2 k^2 \over 2} \right) \right | ^2.
\label{eq:mag-window}
\end{equation}
Substituting Eq.~(\ref{eq:powerlaw}) to Eq.~(\ref{eq:mag-window}),
we can associate $B_\lambda$ with $B_{\rm n}$,
\begin{equation}
B_\lambda ^2 ={B_{\rm n} ^2 \over (2 \pi)^2 \lambda^3 } ({ k_{\rm n} \lambda })^{-n}
\Gamma((n+3)/2).
\label{eq:normalizedmag}
\end{equation}
We take $h^{-1}~$Mpc as $\lambda$ throughout our paper.

Before the recombination epoch, primordial magnetic fields have a
damping scale due to the dissipation of the magnetic fields by the 
interaction between electrons and photons
\citep{jedamzik-katalinic-98,subramanian-barrow-98}. 
As a result, the magnetic
field power spectrum has a sharp cutoff around the damping scale.
The damping scale $1/k_{\rm c}$ after the recombination epoch 
can be related to the magnetic field strength $B_\lambda$ in the
power-law magnetic field case as
\begin{equation}
k_{\rm c} =
\left[ 143 \left ( {B_{\lambda} \over 1 {\rm nG}} \right ) ^{-1}
  \left ( { h  \over 0.7} \right )^{1/2} 
  \left ({ h^2 \Omega_{\rm b} \over 0.021} \right )^{1/2}
\right ]^{2/n+5} {\rm Mpc}^{-1} ,
\label{eq:cutoff}
\end{equation}
in the matter dominated epoch.

Primordial magnetic fields affect the motion of ionized baryon by
the Lorentz force even after the recombination epoch
\citep{wasserman-78,kim-olinto-96}.
Although the residual ionized baryon rate to total baryon is small
after recombination, the interaction between ionized and neutral
baryon is strong in redshifts considered here.
Using the MHD approximation to the baryon fluid, we can write
the evolution equations of density fluctuations with primordial
magnetic fields as, 
\begin{equation}
{\partial^2  \delta_{\rm b} \over \partial t^2} = -2 {\dot a \over a}
{\partial \delta_{\rm b} \over \partial t }
+4 \pi G (\rho _{\rm b} \delta_{\rm b} + \rho _{\rm d} \delta_{\rm d} ) 
+ S(t,{\bm x}),
\label{eq:baryon-den}
\end{equation}
\begin{equation}
S(t,{\bm x})={ \nabla \cdot \left( (\nabla \times {\bm B}_{0} ({\bm x})) 
\times  {\bm B}_{0} ({\bm x}) \right) 
\over 4 \pi \rho_{{\rm b} 0} a^3 (t) },
\end{equation}
\begin{equation}
{\partial^2  \delta_{\rm d} \over \partial t^2} = -2 {\dot a \over a}
{\partial \delta_{\rm d} \over \partial t }
+4 \pi G (\rho _{\rm b} \delta_{\rm b} + \rho _{\rm d} \delta_{\rm d} ),
\label{eq:dm-den}
\end{equation}
where  $\rho_{\rm b}$ and $\rho_{\rm d}$ are the baryon density 
and the dark matter density, and $\delta_{\rm b}$ and 
$\delta_{\rm dm}$ are the density contrast of baryon and dark matter,
respectively. Solving Eqs.~(\ref{eq:baryon-den}) and
(\ref{eq:dm-den}), we can obtain the power spectrum of the density fluctuations.
With the assumption that there is no correlation between primordial
magnetic fields and primordial density fluctuations, 
the density matter power spectrum can be separated into two parts as
\begin{equation}
P(k,t) = P_{{\rm P} }(k,t)+P_{{\rm M} }(k,t), 
\label{matter-power}
\end{equation}
where the first term $P_{{\rm P} }(k)$ is originated from the
primordial density fluctuations,  whose growth rate is proportional to
$t^{2/3}$ in the matter dominated epoch. The second term represents
the power spectrum of the density fluctuations produced by the
primordial magnetic fields.
The power spectrum $P_{{\rm M} }(k)$ is written as 
\begin{equation}
P_{{\rm M} }(k)     = 
\left ( {\Omega _{\rm b} \over \Omega _{\rm m}} \right ) ^2
\left ( t_{\rm i}^2 \over 4 \pi \Omega_{\rm b}  \rho_{c 0}a^3 
(t_{\rm i}) \right)^2
D_{{\rm M} }(t)^2 I^2 (k),
\label{power-mag-part}
\end{equation}
where $\rho_{c0}$ is the critical density at the present epoch,
$t_{\rm i}$ is the initial time which is the recombination epoch
in our calculation,
$D_{{\rm M} }(t)$ is the growth rate and asymptotically proportional to
$t^{2/3}$ in the matter dominated epoch \citep{tashiro-sugiyama-sz}, and
\begin{equation}
I^2 (k) \equiv \langle |\nabla \cdot (\nabla \times {\bm B}_0 ({\bm
x})) \times {\bm B}_0 ({\bm x})|^2 \rangle.
\label{nonline-convo}
\end{equation}
Under the assumption of the isotropic Gaussian statistics for primordial
magnetic fields, we can rewrite the nonlinear convolution
Eq.~({\ref{nonline-convo}}) as 
\citep{wasserman-78, kim-olinto-96}
\begin{equation}
I^2 (k)
= \int dk_1 \int d \mu {P_B (k_1) P_B(|{\bm k} -{\bm k}_1|)
\over (4 \pi)^2 |{\bm k} -{\bm k}_1 |^2 } (2 k^5 k_1^3 \mu+ k^4 k_1 ^4 (1-5 \mu^2) +
2 k^3 k_1^5 \mu^3),  
\label{nonline-convo-2}
\end{equation}
where $\mu$ is $\mu = {\bm k} \cdot {\bm k}_1/ |{\bm k}||{\bm k_1}|$.
Note that the range of integration of $k_1$ in Eq.~(\ref{nonline-convo-2})
depends on $ k$ because we assume that the power spectrum has
a sharp cutoff below $1/k_{\rm c}$ so that $k_1 <k_{\rm c}$ and
$|{\bm k} -{\bm k}_1 |<k_{\rm c}$ must be satisfied.

Eq.~(\ref{nonline-convo-2}) can be estimated analytically 
in the limit of $k/k_ {\rm c} \ll 1$ as 
$I^2(k) \sim \alpha B_\lambda^{2n+10} k^{2n+7} +
\beta B_\lambda^7 k^{4}$
where $\alpha$ and $\beta$ are coefficients which
depend on $n$.  Here we employ the
fact that the damping scale $k_{\rm c}$ is proportional to $B^{-1} _{\lambda}$ as is shown in Eq.~(\ref{eq:cutoff}). 
The former term dominates if
$n < -1.5$, while the latter dominates for $n > -1.5$.  
Accordingly,
if magnetic fields have a power-law index smaller than $-1.5$,
the power-law index of density fluctuations depends on 
that of magnetic fields.
However, if the power-law index of the magnetic fields is
larger than $-1.5$,
the power-law index of the density fluctuations is about 4 
and the amplitude is decided by their damping scale.

We introduce an important scale for the evolution of
density perturbations, i.e., magnetic Jeans length.  
Below this scale, the magnetic pressure gradients, which we do not
take into account in Eq.~(\ref{eq:baryon-den}), 
counteract the gravitational force 
and prevent further evolution of density fluctuations. 
The magnetic Jeans scale is evaluated as \citep{kim-olinto-96} 
\begin{equation}
k_{\rm MJ}    
= \left[ 13.8 \left ( {B_{\lambda} \over 1 {\rm nG}} \right ) ^{-1}
\left ( { h^2 \Omega _{\rm m} \over 0.18} \right )^{1/2} 
\right ]^{2/n+5} {\rm Mpc}^{-1} .
\label{jeans}
\end{equation}
For simplicity,
we assume that the density fluctuations do not grow
below the scale,
although the density fluctuations below the scale are, in fact, oscillating like the baryon oscillation.

\section{Mass function and X-ray observation}

The additional density fluctuations produced by primordial magnetic
fields enhance the number of dark matter halos. In order to estimate
this enhancement, we use the mass function which is calibrated to 
fit the numerical simulation by \citet{tinker-kravtsov-08},
\begin{equation}
{dn \over dM} = A{\Omega_{\rm m} \rho_{c0} \over M} {d\ln \sigma^{-1}
\over dM}\left[\left( {\sigma \over b}\right)^{-a} + 1\right]
e^{-c/\sigma^2},
\end{equation}
\begin{equation}
A(z) = A_0 \left(1+z\right)^{-0.14},~
a(z) = a_0 \left(1+z\right)^{-0.06},~
b(z) = b_0 \left(1+z\right)^{-\alpha},~
\log \alpha(\Delta) = -\left(\frac{0.75}{\log(\Delta /75)}\right)
^{1.2},
\end{equation}
where $\sigma$ is the smoothed variance of the density fluctuations
with a top-hat window function,
$A_0$, $a_0$, $b_0$ and $c$ depend on $\Delta$, and $\Delta$ is the
overdensity contrast within a sphere of radius $R_\Delta$ which is
related to the halo mass $M_\Delta$ by
\begin{equation}
M_\Delta= \Omega_{\rm m} {4 \pi \over 3} R_\Delta ^3 \rho_c \Delta.
\end{equation}
For example, in the case of the halo virial mass, $\Delta$ is
$\Delta=178$ in the matter dominated epoch.

Our interest is to examine the potential of the X-ray galaxy 
cluster observation to give constraints on primordial magnetic
fields through the halo mass function. 
In the X-ray observation, the halo mass for high $\Delta$ is more
robust than for low $\Delta$. Therefore, according to 
\citet{vikhlinin-chandora-09}, we set $\Delta=500$.
The observational luminosity threshold gives the mass threshold 
for observed halos. Therefore, the number count of halos over 
the luminosity threshold corresponds to the integrated mass function,
\begin{equation}
N(>M_L)=\int_{M_L} d M~{dn  \over dM}, 
\end{equation}
where $M_L$ is the mass threshold corresponding to the luminosity
threshold.

\begin{figure}
  \begin{tabular}{cc}
   \begin{minipage}{0.5\textwidth}
  \begin{center}
\includegraphics[keepaspectratio=true,height=70mm]{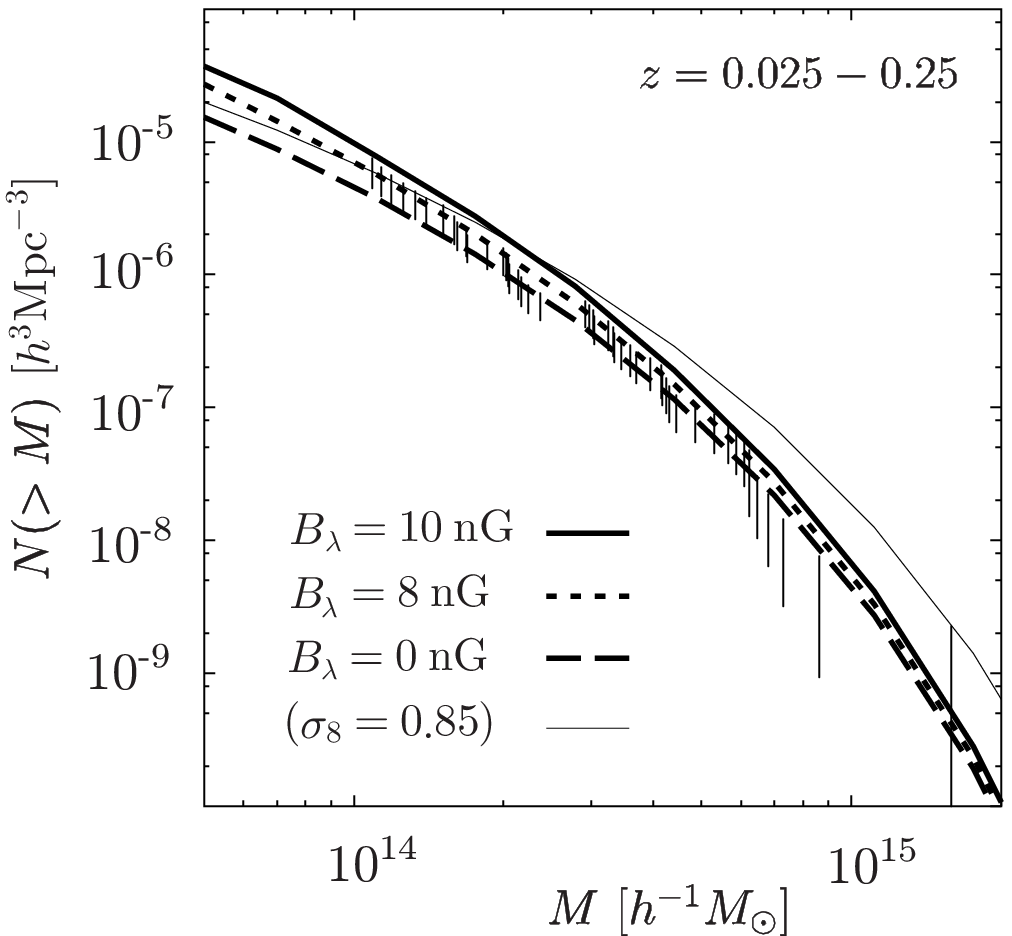}
  \end{center}
  \end{minipage}
   \begin{minipage}{0.5\textwidth}
  \begin{center}
\includegraphics[keepaspectratio=true,height=70mm]{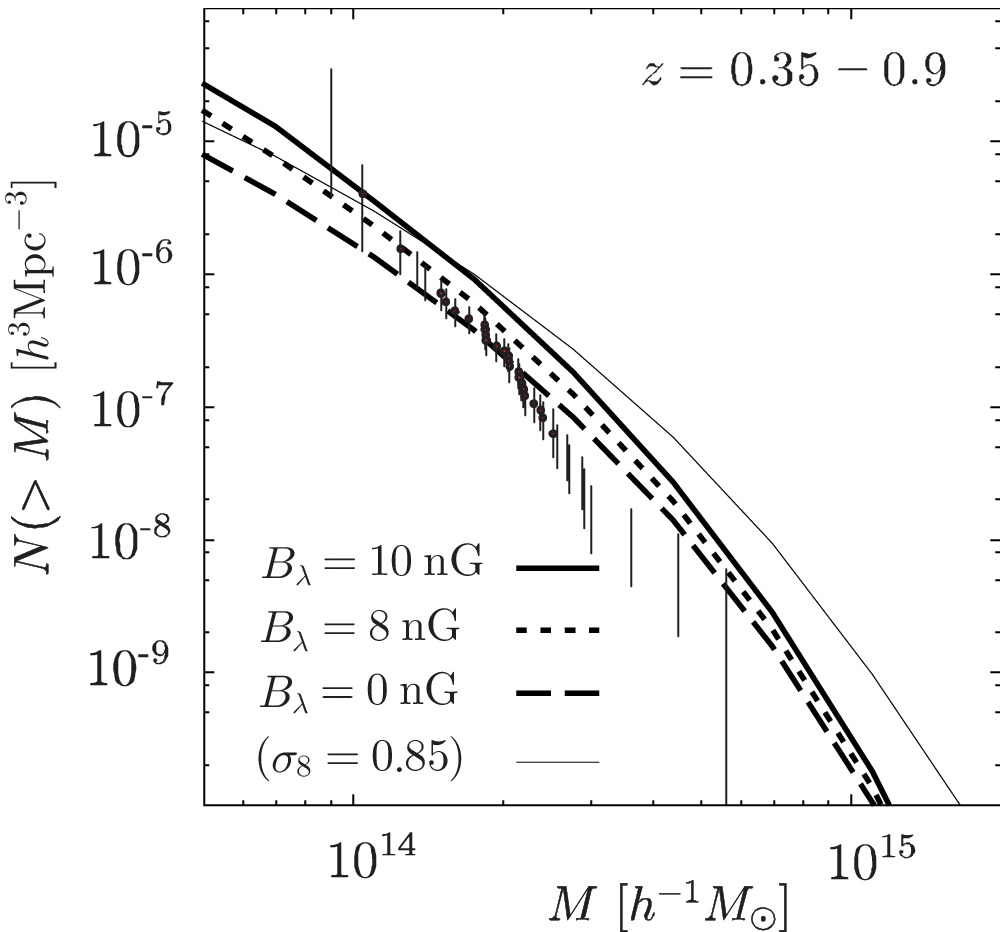}
  \end{center}
   \end{minipage}
  \end{tabular}
  \caption{
  The integrated number counts of X-ray galaxy clusters. 
  The left panel
  shows the number count for the low redshift bin $z=0.025-0.25$, 
  while the right panels shows that for the high redshift bins
  $z=0.35-0.90$. In both panels, the solid, the dashed and the dotted
  lines represent the number counts for the case of $B_\lambda=10~$nG,
  $B_\lambda=8~$nG and $B_\lambda=0~$nG, respectively. 
  We also plot the results of {\it Chandra} 
  in \citet{vikhlinin-chandora-09}.
  For comparison, we put the
  number counts for the $B_\lambda=0~$nG case with $\sigma_8=0.85$
  as the thin solid
  line.
  }
  \label{fig:numbercount}
\end{figure}

In Fig.~\ref{fig:numbercount}, we plot the integrated number count 
as a function of the mass threshold for each primordial magnetic 
field strength with the magnetic field spectral index $n=-2.8$.
The left panel in Fig.~\ref{fig:numbercount} is for the low redshift
bin, $z=0.025$--$0.25$, while the right panel is for the high
redshift bin, $z=0.35$--$0.9$.
Although the additional density power spectrum induced by primordial magnetic fields dominate the primordial power spectrum on smaller
scales than $1~$Mpc, the additional power spectrum can enhances 
$\sigma_8$ to 0.8 for $B_\lambda=10~$nG. 
For reference, we put the integrated number count
for the $\Lambda$CDM model with $B_\lambda=0~$nG and $\sigma_8=0.85$.
The existence of the primordial magnetic fields lift up the
mass function on small scales, because
the additional density fluctuations produced by primordial magnetic
fields have a blue spectrum. As a result, compared with the case 
with $B_\lambda=0~$nG and $\sigma_8=0.85$, while the number 
count in the case with primordial magnetic fields is small on large
scales, it is more enhanced on small scales.

\citet{vikhlinin-chandora-09} obtained galaxy cluster mass functions
at two redshift range using {\it Chandra} observation data and
concluded that the obtained mass functions are in good agreement
with the cosmological model with $\sigma_8=0.75$. We compare the theoretical
mass functions with the data with the error bars due to the Poisson
uncertainties in Fig.~\ref{fig:numbercount}.
From this figure, we can conclude that $Chandra$ observations rule out
the primordial magnetic field strength, $B_\lambda  \gtrsim 8~$nG
at roughly one-sigma.

\section{S-Z number counts}

The S-Z effect is caused by the scattering of CMB photons with
electrons in hot gas in galaxy clusters.
The change of the CMB intensity with the frequency $\nu$
by the S-Z effect is expressed, in the R-J limit, as
\citep{szeffect-1972,birkinshaw-sz-99}
\begin{equation}
I_\nu (\theta) = 2 \nu^2 T_{\rm CMB} g(x) y(\theta),
\end{equation}
where $T_{CMB}$ is the CMB temperature and $g(x)$ is the S-Z 
effect spectral shape given by $g(x) = x^2 e^x (x/\tanh(x/2)-4) /(e^x-1)^2$
with $x=2 \pi \nu /T_{\rm CMB}$.
The Compton $y$-parameter is given by the integral of the electron 
gas pressure along the line of sight at $\theta$
\begin{equation}
y(\theta)=\int dl ~f_g{T_e \over \mu_e m_e} n_e \sigma_T,
\label{eq:ypara}
\end{equation}
where $T_e$ is the electron gas temperature, $n_e$ is the electron
density, $\sigma_{\rm T}$ is the Thomson
scattering cross-section, $m_{e}$ is the electron mass,
$\mu_e$ is the mean mass per electron $\mu_e=1.143$ 
and $f_g$ is a gas fraction in a galaxy cluster which we set
$f_g=0.12$ \citep{mohr-mahiesen-99}.   
In the S-Z cluster survey, it is assumed that the S-Z cluster is a
point like source within the telescope beam. Therefore, we consider the
total flux density from a cluster at redshift $z$ by integrating the 
cluster surface,
\begin{equation}
S_\nu = 2 \nu^2 T_{\rm CMB} g(x) {Y \over D_a^2},
\end{equation}
where $D_a$ is the angular diameter distance to the cluster at $z$
and $Y$ is the integrated $y$-parameter over the cluster surface,
\begin{equation}
Y= \int d \Omega ~y(\theta).
\label{eq:large-Y}
\end{equation}
In order to calculate Eq.~(\ref{eq:large-Y}), we need the electron
density profile in a galaxy cluster. We take the assumption that
the electron density profile is isothermal $\beta$-profile 
\citep{cabariere-beta-1976},
\begin{equation}
n_{e}(r)=\left \{
\begin{array}{cc}
n_0 \left(1+ {r^2 \over R_{\rm c}^2} \right)^{-{3\beta \over 2}}
&
r<R_{\rm v} \\
0 & r\ge R_{\rm v}
\end{array}
\right.
\end{equation}
where $R_c$ is the core radius of galaxy cluster which is related to
the virial radius $R_{\rm v}$ with the parameter $s(z)$ as 
$R_c = R_{\rm v}/s(z)$.
Following \citet{2002MNRAS.336.1256K},
we set
\begin{equation}
s(z) \approx \frac{10}{1+z}\left[\frac{M}{M_*(0)}\right]^{-0.2},
\label{eq:concentrait}
\end{equation}
where $M_*(0)$ is a solution to
$\sigma(M)=\delta_c$  at the redshift $z=0$, where 
$\delta_c$ is the critical density contrast for collapsing.

Taking the assumption that the galaxy clusters are spherical and in 
the hydrodynamical equilibrium, we can relate the 
virial mass $M_{\rm v}$ with the virial radius,
\begin{equation}
M_{\rm v}={4\pi\over 3} \rho_{\rm M}(z)\Delta_{\rm v}(z)R_{\rm v}^3,
\end{equation}
where $\Delta_{\rm v}(z)$ is the overdensity contrast for
virialization \citep{nakamura-suto-97}. 

We introduce the electron density weighted average temperature
$\langle T_{\rm e}\rangle_n$, which is defined by $\langle T_{\rm e}
\rangle_n \equiv \int d l n_{e} T_e / \int dl n_e$.
\citet{battye-weller-03} has obtained $\langle T_{\rm e}\rangle_n$
under the isothermal assumption in the $\lambda$CDM model,
\begin{equation}
\langle T_{\rm e}\rangle_n= T_\star ~{\rm keV}
\left[{M_{\rm vir}\over 10^{15}h^{-1}M_{\odot}} \right]^{2/3}
\left[\Delta_{\rm c}(z) {H(z)^2 \over H_0^2}\right]^{1/3}
\left[1-2{(1-\Omega_{m}(z))\over \Delta_{\rm c}(z)}\right],
\label{eq:ele-temp}
\end{equation}
where $T_\star$ is the temperature normalization factor and 
they adopt $T_\star=1.6$ for agreement with numerical 
simulation works in \citet{bryan-norman-98} and 
\citet{pierpaoli-scott-01}.

Using the $\beta$-profile assumption with the electron density 
weighted average temperature given by in Eq.~(\ref{eq:ele-temp}), 
we can write the $y$-parameter as
\begin{equation}
y(\theta)= {\langle T_{e}\rangle _{n}\over
m_{\rm e}}{f_{\rm gas}M_{\rm vir}\over \mu_{e}m_{\rm p}} 
\sigma_{\rm T} \zeta(\theta),
\end{equation}
where $\zeta(\theta)$ is the projected profile of the electron
density,
\begin{equation}
\zeta(\theta) =
\left(1+{\theta^2 \over \theta_c^2}\right)^{-{1 \over 2}}
{\tan^{-1} \left[ \left(
{s^2-\theta^2/\theta_c^2 \over 1+\theta^2/\theta_c^2}
\right)^{1/2} \right] \over{\tan^{-1} s}},
\end{equation}

In actual observations, the finite beam size of telescopes causes
the beam-smearing effect. This effect can be accounted by modifying
Eq.~(\ref{eq:large-Y}) to \citep{bartlett-00}
\begin{equation}
Y= \int d \Omega~ y(\theta) B(\theta).
\end{equation}
Here we assume that the beam profile is described in a Gaussian form
$B(\theta)=\exp[-\theta^2/(2\sigma_b^2)]$ with $\sigma_b =
\theta_{\rm FWHM}/\sqrt{8\ln 2}$ where $\theta_{\rm FWHM}$ is the 
full-width-half-maximum (FWHM).

\begin{figure}
  \begin{center}
    \includegraphics[keepaspectratio=true,height=70mm]{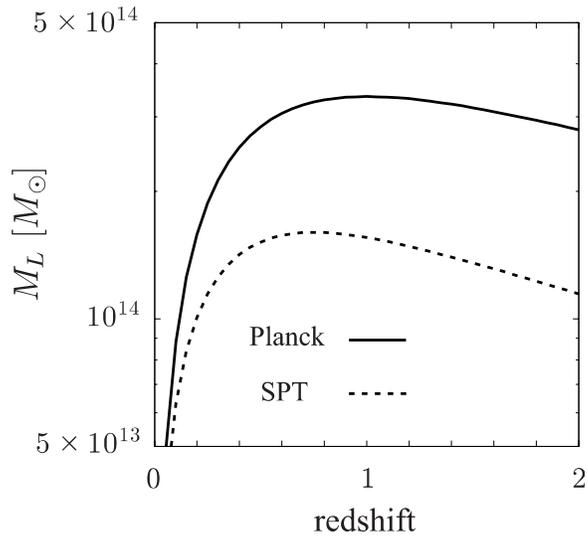}
  \end{center}
  \caption{The limiting mass for each redshift. The solid line 
  is for {\it Planck} and the dashed line is for SPT. The each
  experimental parameters are in Table.~\ref{table:ex-para}. }
  \label{fig:mlim}
\end{figure}

The parameter $Y$ depends on the mass and the redshift of galaxy
clusters. Therefore, giving the flux limit $S_{\rm lim}$ of the 
observation, we can obtain the limit mass of galaxy clusters $M_L$
at each redshift. We show
$M_L$ for {\it Planck} and SPT in Fig.~\ref{fig:mlim}. We present the
parameter value for each observation in Table~\ref{table:ex-para}.
{\it Planck} will cover the full sky and the {\it Planck}
sensitivity is 14 mJy at 100 GHz. SPT covers $\Delta \Omega=4000$
degree square and the SPT sensitivity is 0.8 mJy at 150 GHz.
SPT has a better sensitivity than {\it Planck}.
As a result, $M_L$ for SPT is lower than for {\it Planck}.

\begin{figure}
  \begin{tabular}{cc}
   \begin{minipage}{0.5\textwidth}
  \begin{center}
\includegraphics[keepaspectratio=true,height=70mm]{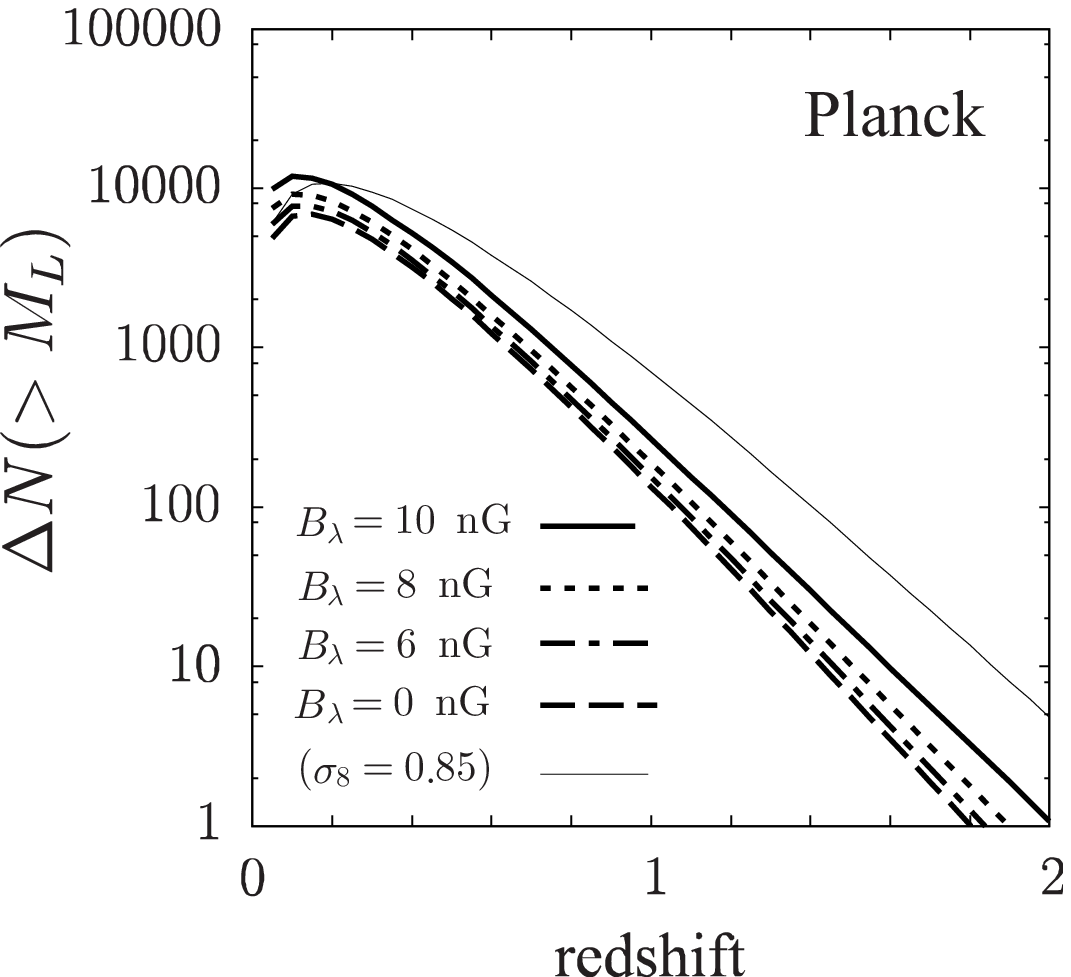}
  \end{center}
  \end{minipage}
   \begin{minipage}{0.5\textwidth}
  \begin{center}
\includegraphics[keepaspectratio=true,height=70mm]{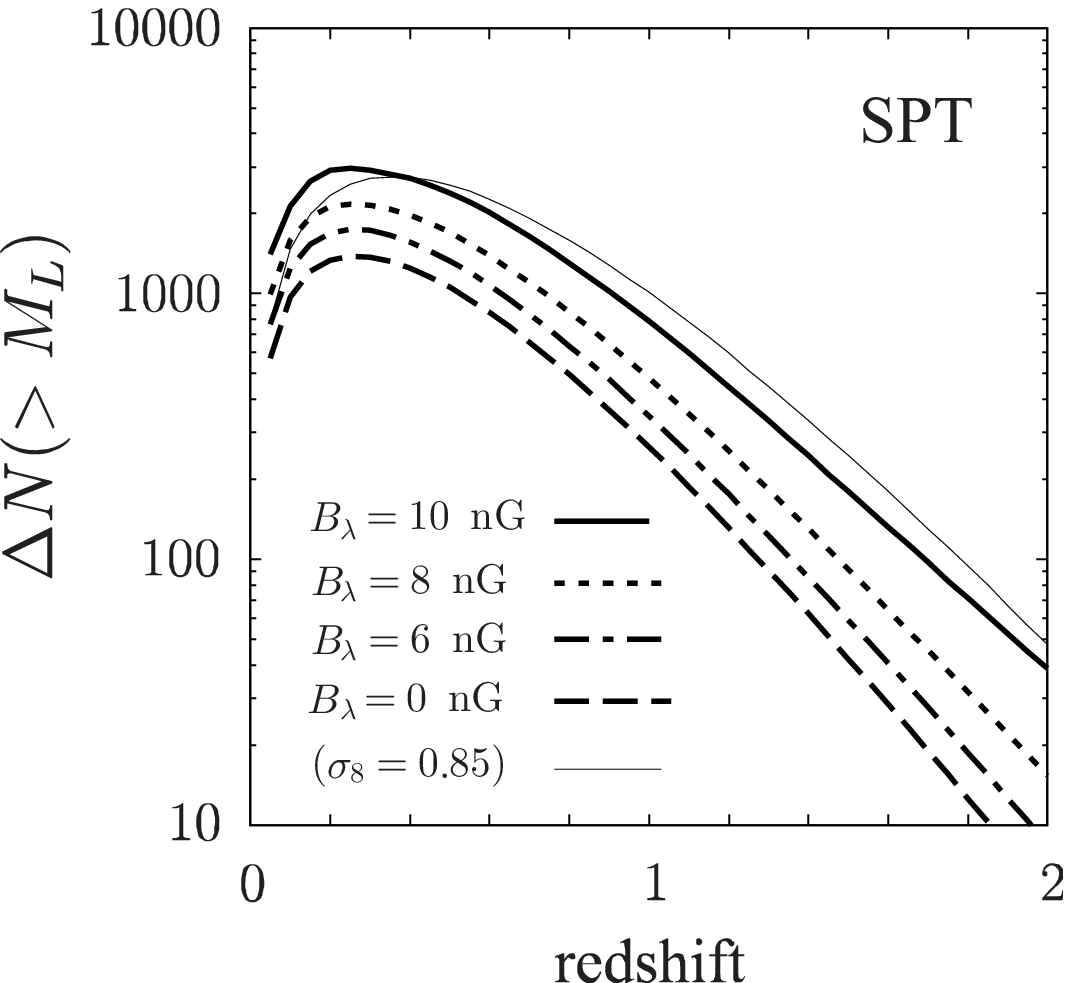}
  \end{center}
   \end{minipage}
  \end{tabular}
  \caption{
  The integrated number counts of S-Z galaxy clusters. 
  The left panel is for {\it Planck}, and the right panels is  for 
  SPT. In both panels, the solid, the dashed, 
  the dashed-dotted and the dotted lines represent the number counts
  for the case of $B_\lambda=10~$nG, 
  $B_\lambda=8~$nG, $B_\lambda=6~$nG and $B_\lambda=0~$nG, respectively. 
  For comparison, we put the
  number counts for the $B_\lambda=0~$nG case with $\sigma_8=0.85$ as the thin solid line.
  }
  \label{fig:SZnumber}
\end{figure}

The combination between the S-Z galaxy cluster survey and the
follow-up optical observation enables us to obtain the cluster number
count for redshift bins.
In Fig.~\ref{fig:SZnumber},
we show the number count of galaxy clusters with mass higher than
the limiting mass shown in Fig.~\ref{fig:mlim},
\begin{equation}
\Delta N(z)= {dn \over dz} \Delta z =\Delta \Omega  \int_{M_{L}} dM 
{dV \over dz d\Omega}  {dn \over dM } {\Delta z}  .
\end{equation}
In both panels of Fig.~\ref{fig:SZnumber}, we set $\Delta z=0.1$
and the primordial magnetic field spectral index $n=-2.8$.

Fig.~\ref{fig:SZnumber} shows that the predicted number counts of
{\it Planck} and SPT are almost same in low redshifts. 
This is because, although $\it Planck$ has less sensitive to small
galaxy clusters  than SPT, $\it Planck$'s full sky survey area
increases the number of the observed galaxy clusters, and vice versa.
However, in high redshift, since the number of large-mass clusters
rapidly decreases, the number count for $\it Planck$ become much
lower than for SPT.

Primordial magnetic fields generate additional density fluctuations
in small scales and bring the early structure formation.
Therefore, the difference from $\Lambda$CDM cosmology due to
the existence of primordial magnetic fields is 
emphasized in the SPT observation, especially in high redshifts.
Even the primordial magnetic fields with $B_\lambda=6~$nG amplifies
the number count in high redshifts by 50 \% for the SPT sensitivity,
while, for {\it Planck}, such primordial magnetic fields cannot bring
a significant amplification.

For reference, we put the number counts for the $B_\lambda=0~$nG 
case with $\sigma_8=0.85$ as the thin solid line in both panels in 
Fig.~\ref{fig:SZnumber}. 
Although $\sigma_8$ in the case of $B_\lambda=10~$nG
corresponds to $0.8$, the power spectrum in the case
of $B_\lambda=10~$nG has larger amplitudes in high $k$s than
in the case of $\sigma_8=0.85$ without primordial magnetic fields.
This results in the fact the number of galaxy clusters with small mass
is larger in the case of $B_\lambda=10~$nG than in the case of
$\sigma_8=0.85$. Therefore, 
the number counts for $B_\lambda=10~$nG exceed the ones for 
$\sigma_8=0.85$ in the low redshifts where
both {\it Planck} and SPT are sensitive to low mass clusters
as shown Fig.~\ref{fig:mlim}.
In particular, the number counts of SPT for $B_\lambda=10$nG is 
almost same as for $\sigma_8=0.85$ even in high redshifts, because SPT has small limiting mass.

\begin{table}
\centering
\begin{tabular}{ccccc}
\hline 
\hline 
 & $S_{\rm lim} $ & $\nu~[{\rm GHz}]$ & $\theta_{\rm
fwhm}$ & $\Delta\Omega~ [{\rm deg}^2]$ \\
\hline 
\hline 
Planck & $14~$mJy & $100$ & $9.^\prime$ & full sky \\
\hline 
SPT& $0.8~$mJy & $150$ & $1.^\prime$ & $4000$ \\
\hline 
\hline 
\end{tabular}
\caption{The experimental parameters for S-Z surveys.}
\label{table:ex-para}
\end{table}

\section{conclusion}

In this paper, we have studied the effect of primordial magnetic
fields on the galaxy survey by X-ray and S-Z observation.
The primordial magnetic fields generate additional density
fluctuations which has a blue power spectrum. Therefore, the number
of galaxy clusters, especially small ones, is enhanced.
X-ray and S-Z survey can directly observe this enhancement.
Nano-Gauss primordial magnetic fields bring observable enhancement
of the number count by the order of factors.

For X-ray cluster surveys, we have used {\it Chandra}'s result to
put a constraint on the amplitude of primordial magnetic fields.
We have found that {\it Chandra}'s result rules out the existence of
primordial magnetic fields with $B_\lambda  \gtrsim 10~$nG at roughly
one-sigma level.

S-Z cluster surveys also have a sensitivity to constrain primordial
magnetic fields. Especially the observation like SPT which has small
limiting mass with 1 arcmin angular resolution is a good probe of
primordial magnetic fields. We have found that the combination with
high redshift optical surveys has the potential to put the constraint
on the fields of nano Gauss order.

In this paper, we consider only primordial magnetic fields with 
$n=-2.8$. The power spectrum of the density fluctuations
generated by primordial magnetic fields has a dependence on 
the spectral index of the primordial magnetic fields. 
The large spectral index induces the large amplification of the 
density fluctuations on small scales and increases
the mass function for small-mass clusters. 
For example, $B_\lambda=4~$nG and $n=-2.5$ amplifies
the number count in high redshifts by 50 \% for the SPT sensitivity,
comparing with the number count without primordial magnetic fields.
This amplification is same as in the case of $B_\lambda=6~$nG with
$n=-2.5$.
Therefore, SPT has the potential to 
put the strong constraint on the primordial magnetic fields
with large $n$.

In our calculation, we ignore the effect of primordial magnetic
field on the structure of a halo. However, in order to obtain a 
highly accurate constraint on primordial magnetic fields, it is
necessary to study the modification on the electron density
profile and the relation between the X-ray temperature and the 
cluster mass by primordial magnetic fields. For example,
\citet{zhang-2004-sz} and \citet{gopal-roychowdhury-2010} pointed
out that magnetic fields with several $\mu$Gauss in a halo modify the electron density profile and this modification change the S-Z effect
signal. The adiabatic contraction in the halo formation easily 
amplifies the order of nano Gauss of primordial magnetic field 
strength to the order of $\mu$ Gauss.
Taking into account such effects, we will study the constraint on
primordial magnetic fields through X-ray and S-Z surveys in the
future.

\section*{Acknowledgements}
We thank the reviewer for his/her thorough review and highly appreciate the comments and suggestions, which significantly contributed to improving the quality of the publication. HT is supported by the Belgian Federal Office for Scientific,
Technical and Cultural Affairs through the Interuniversity Attraction Pole P6/11.

\end{document}